\documentclass[]{interact}

\usepackage{epstopdf}
\usepackage{subfigure}
\usepackage{color}

\usepackage[numbers,sort&compress,merge]{natbib}
\bibpunct[, ]{[}{]}{,}{n}{,}{,}
\renewcommand\bibfont{\fontsize{10}{12}\selectfont}

\theoremstyle{plain}

\theoremstyle{definition}

\theoremstyle{remark}

\begin{document}

\title{Influence of broken-pair excitations on the exact pair wavefunction}

\author{
\name{Jacob M. Wahlen-Strothman\textsuperscript{a}, Thomas M. Henderson\textsuperscript{a,b}, Gustavo E. Scuseria\textsuperscript{a,b}\thanks{CONTACT Gustavo E. Scuseria. Email: guscus@rice.edu}}
\affil{\textsuperscript{a}Department of Physics  Astronomy, Rice University, Houston, TX, USA; \textsuperscript{b}Department of Chemistry, Rice University, Houston, TX, USA}
}

\maketitle

\begin{abstract}
Doubly occupied configuration interaction (DOCI), the exact diagonalization of the Hamiltonian in the paired (seniority zero) sector of the Hilbert space, is a combinatorial cost wave function that can be very efficiently approximated by pair coupled cluster doubles (pCCD) at mean-field computational cost. As such, it is a very interesting candidate as a starting point for building the full configuration interaction (FCI) ground state eigenfunction belonging to all (not just paired) seniority sectors. The true seniority zero sector of FCI (referred to here as FCI${}_0$) includes the effect of coupling between all seniority sectors rather than just seniority zero, and is, in principle, different from DOCI.  We here study the accuracy with which DOCI approximates FCI${}_0$. Using a set of small Hubbard lattices, where FCI is possible, we show that DOCI $\sim$ FCI${}_0$ under weak correlation.  However, in the strong correlation regime, the nature of the FCI${}_0$ wavefunction can change significantly, rendering DOCI and pCCD a less than ideal starting point for approximating FCI.
\end{abstract}


\section{Introduction}
In an ideal world, we would be able to describe both weak and strong electronic correlations with equal reliability and accuracy, and without having to make any assumptions about the nature of the correlations relevant for a particular problem.  For weakly correlated systems in the ground electronic state, this is (more or less) possible, but strong correlations remain challenging.

One technique which has shown significant potential for accurate treatment of strong correlations is what we will refer to as pair coupled cluster doubles (pCCD)~\cite{Limacher13,Tecmer14,Stein14,Boguslawski14,Henderson14_2,Henderson15,Shepherd16}, in which the wave function is written as the exponential of a double excitation operator which excites both electrons from the same occupied orbital to the same virtual orbital; in other words, the exponential of a cluster operator which preserves seniority, where the seniority of a determinant is the number of singly-occupied spatial orbitals.  While this form of wave function seems rather primitive, pCCD often yields surprisingly accurate energies in strongly correlated systems.  The basic reason that pCCD works as well as it does is that it very accurately approximates the doubly-occupied configuration interaction (DOCI) \cite{Allen62,Smith65,Weinhold67,Veillard67,Couty97,Kollmar03,Bytautas11,Alcoba14,Alcoba14_2} energy and wave function, where DOCI corresponds to diagonalizing the Hamiltonian in the space of all states in which electrons are paired.  Since DOCI generally offers a fairly accurate description of strong correlations and pCCD generally corresponds closely to DOCI, pCCD is also a useful tool for the treatment of strong correlations.  Better yet, while DOCI is significantly cheaper than full configuration interaction (FCI), its cost still scales combinatorially with system size, but pCCD has mean-field computational scaling.

While pCCD and DOCI have significant promise, they also have a crucial weakness in that neither describes any correlations which break electron pairs (i.e. correlations which change seniority).  These residual correlations are generally weaker, but are also very numerous, and they may have a considerable effect on the total energy and wave function.  This effect must be captured in some way if we are to achieve predictive accuracy.  Several attempts have been made to incorporate these correlations within the basic framework of pCCD \cite{Limacher14,Garza15,Garza15b,Boguslawski15}.

Perhaps the simplest approach is frozen-pair coupled cluster doubles (fpCCD) \cite{Stein14} in which the seniority-preserving sector of a coupled cluster doubles (CCD) excitation operator is taken from pCCD and the rest is obtained from the usual CCD equations.  But while fpCCD is more robust in the presence of strong correlations than is CCD, it still fails when the correlations are strong enough.

In this manuscript, we seek to address why this might be the case.  Our basic hypothesis is that while pCCD accurately approximates the DOCI wave function and energy, it need not be the case that DOCI accurately approximates the zero-seniority portion of the exact FCI wave function.  In other words, because the seniority sectors of the Hamiltonian are not decoupled in general, it is quite possible that the DOCI wavefunction may become incorrect in certain physically relevant limits.  If this happens -- if, that is, the coupling between seniority sectors is strong -- then it is not \textit{a priori} clear that DOCI or pCCD form a suitable starting point for including the effects of the rest of the Hamiltonian.

\section{The Hubbard Model}
To test our hypothesis, we will perform benchmark calculations on the Hubbard model Hamiltonian,
\begin{equation}
H = -t\sum_{\langle ij\rangle\sigma}a^\dagger_{i\sigma}a_{j\sigma} + U\sum_{i}n_{i\uparrow}n_{i\downarrow},
\end{equation}
where $\langle ij\rangle$ represents nearest-neighbor lattice sites, $t$ is a kinetic hopping term, and $U$ is the on-site repulsion. This is a highly studied~\cite{Shi13,Shi14,Fano90,LeBlanc15} and useful testing ground for a number of reasons. It has already been used to study and test DOCI, pCCD, and other seniority-based coupled cluster methods~\cite{Stein14,Henderson14,Shepherd16}. The form of the ground state solution can be dominated by weak correlation when $U/t$ is small, or strong correlation when large.  The Hubbard model thus allows us to tune the correlation strength by changing a single parameter.  Restricted Hartree-Fock (RHF) significantly overestimates the on-site repulsion for $U>0$, and produces very poor energies for any value of $U/t$ that is not near zero. Coupled cluster doubles (CCD)~\cite{Bartlett07} significantly overcorrelates as $U$ increases and will eventually result in complex solutions in 1D. This failure also occurs in nearly all cases for half-filled 2D systems even when $U$ is small. In addition there are significant contributions from both weak and strong correlation in the intermediate coupling regime. This results in a system in which it is very difficult to get accurate energies as methods focusing only on strong or weak correlations separately will fail at some point.

This Hamiltonian has the added value as a benchmark when we consider doped systems. A Hubbard lattice with holes at large values of $U/t$ contains strong competition between delocalised electrons and antiferromagnetic structure requiring an effective description of both weak and strong correlations for accurate results. The onset of strong correlation is typically marked by the presence of symmetry breaking in the Hartree-Fock wavefunction at a critical value of the interaction ($U_c$). For all the Hubbard systems we consider in this work, there is an RHF to UHF instability in the mean field for small values of $U$ ($U_c<2t$ in all cases). For the purposes of later notation, we will work in units where $t=1$.

\section{DOCI and pCCD}

As described above, DOCI is exact diagonalization in the subspace of all pair excitations,
\begin{equation}
H|DOCI\rangle = E_0|DOCI\rangle + |\Omega\neq 0\rangle,
\end{equation}
where $\Omega$ is the seniority of the wavefunction representing the number of broken electron pairs. We only consider states that can be built from the mean-field ground state with pair excitations of the form,
\begin{equation}
P^\dagger_aP_i = c^\dagger_{a\uparrow}c^\dagger_{a\downarrow}c_{i\downarrow}c_{i\uparrow},
\end{equation}
in order to preserve the seniority.
The pairs, and by extension the seniority, is defined by the basis of the reference wavefunction. Therefore, the basis in which we make the comparison is important, and we will see how the basis affects the results.

The usefulness of this wavefunction stems from the observation that the DOCI factorises into the exponential ansatz of paired coupled cluster doubles for many systems \cite{Shepherd16},
\begin{equation}
|pCCD\rangle = e^{T_p}|\Phi\rangle,\quad T_p=\sum_{ai}t_{i}^{a}P^\dagger_aP_i .
\end{equation}
This has been observed for repulsive systems, but does not hold for attractive systems such as the attractive reduced pairing model~\cite{Degroote16}. As seniority is a symmetry of this system, DOCI is equivalent to FCI, but pCCD fails outside the weakly correlated limit. This is because it is an exponential ansatz and cannot produce the Bessel form of the projected BCS wavefunction, which is the correct solution for the strongly correlated reduced pairing model. If we stick to repulsive systems, such as the repulsive Hubbard model, pCCD is insulated from the breakdown of traditional coupled cluster under strong correlations, as it very accurately approximates the results of DOCI which is a variational method and cannot overcorrelate.

\begin{figure*}[b]
\includegraphics[width=0.99\textwidth]{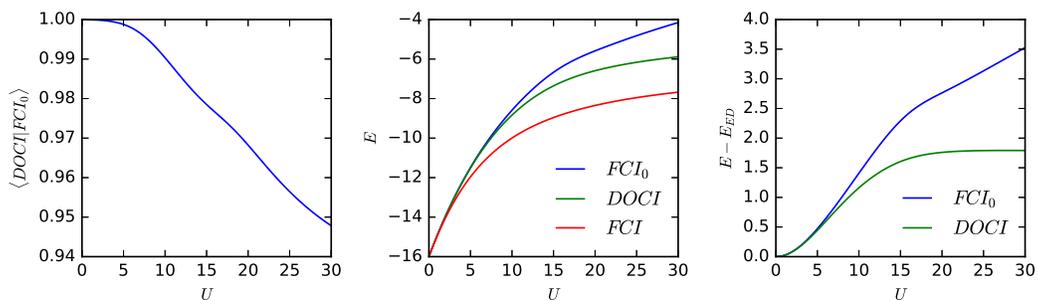}
\caption{\label{fig:ov_pCCD_PBC} The wavefunction overlap and energies for the DOCI and FCI${}_0$ wavefunctions for a $2\times 4$ periodic Hubbard lattice with 6 particles in the oo-pCCD basis. }
\end{figure*}

\begin{figure*}[ht!]
\includegraphics[width=0.99\textwidth]{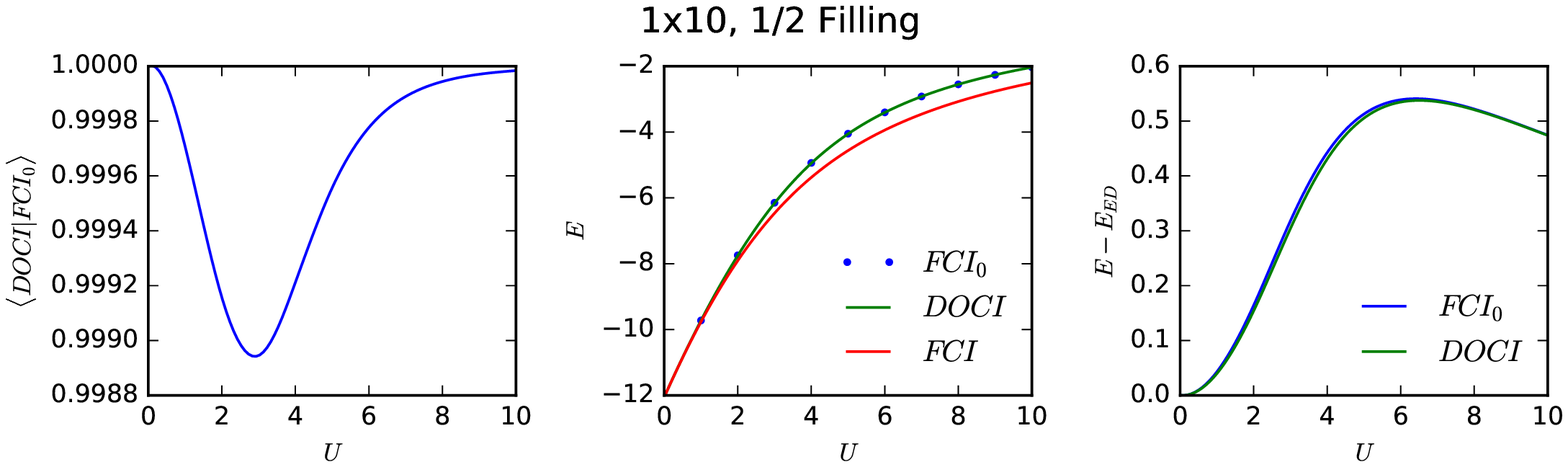}\\
\includegraphics[width=0.99\textwidth]{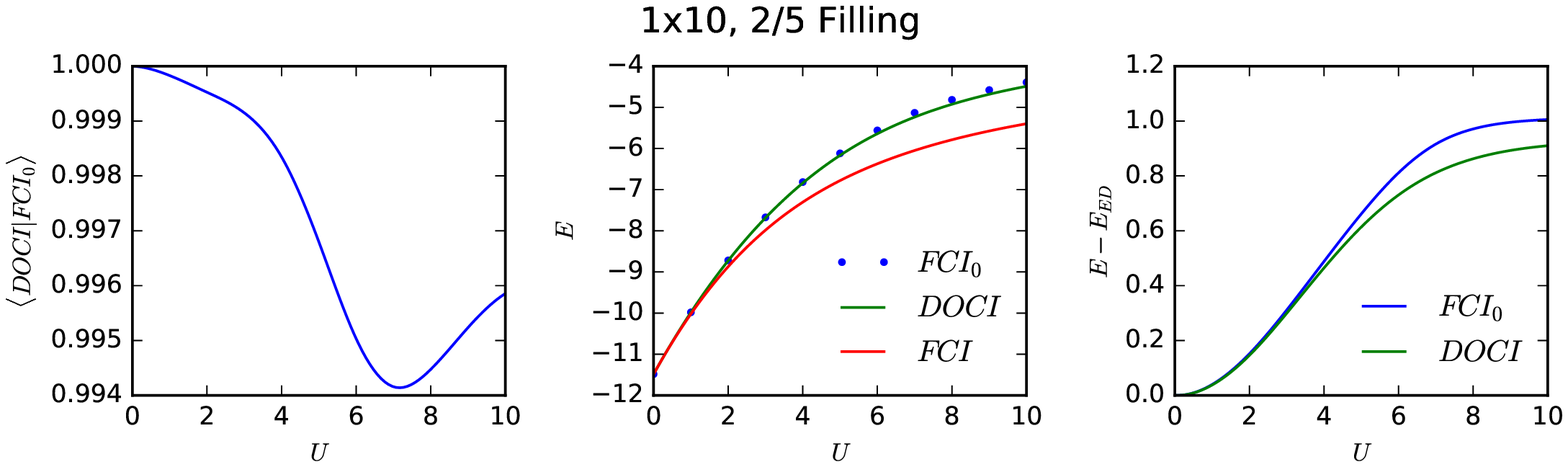}\\
\includegraphics[width=0.99\textwidth]{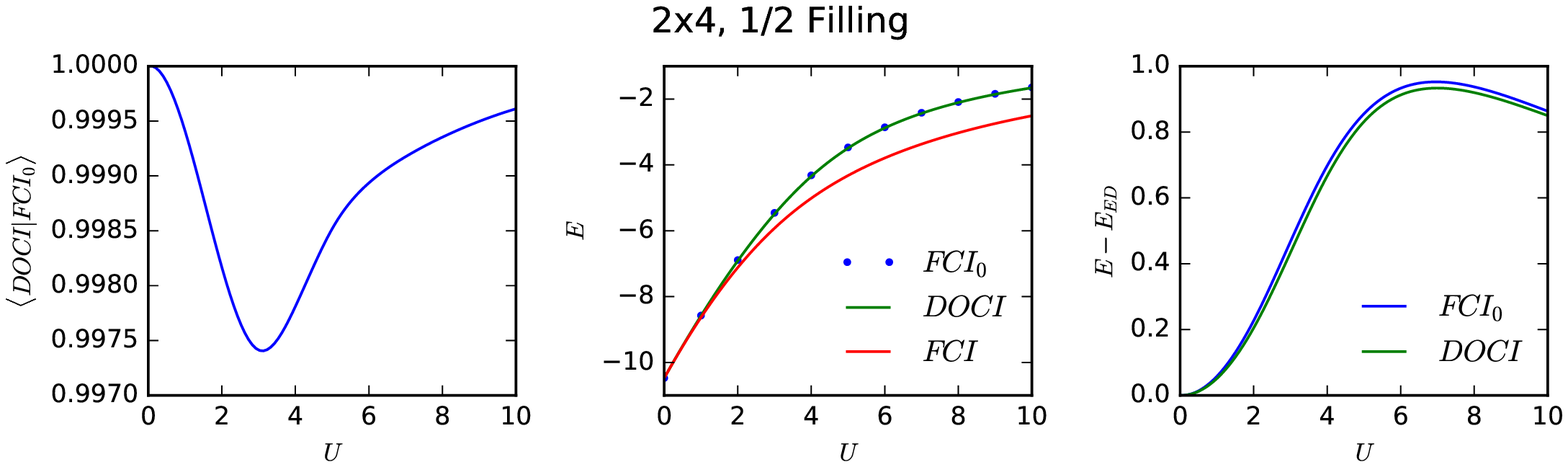}\\
\includegraphics[width=0.99\textwidth]{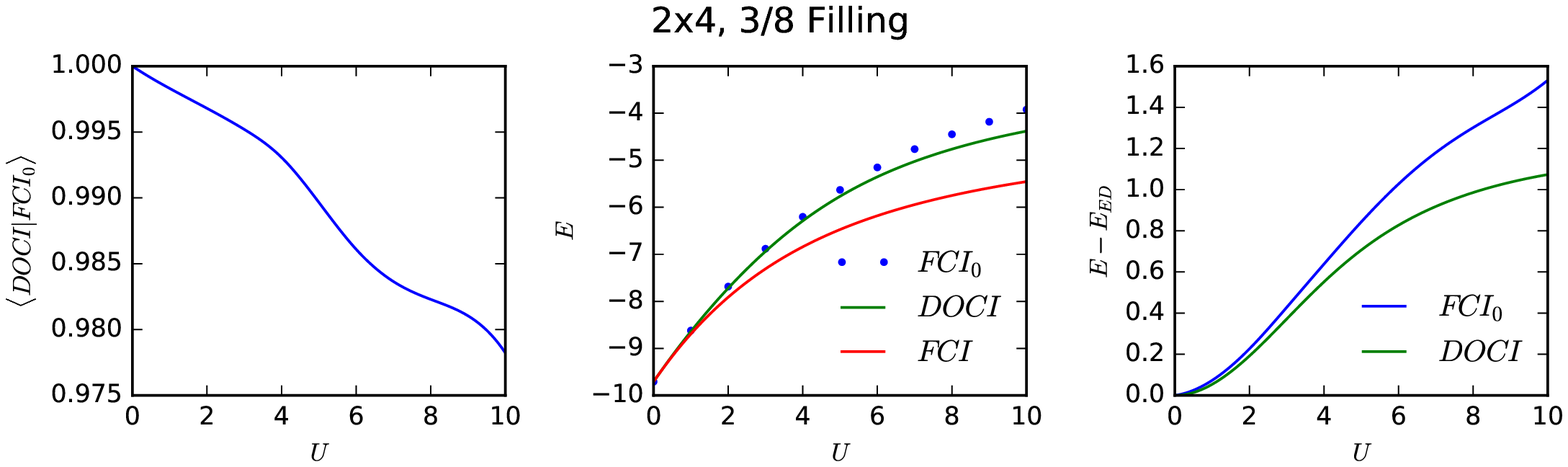}
\caption{\label{fig:ov_pCCD} The overlap and energies of the DOCI and FCI${}_0$ wavefunctions for non-periodic Hubbard chains ($1\times 10$) and ladders ($2\times 4$) in the oo-pCCD basis.}
\end{figure*}

\section{Results}

In order to test the DOCI wavefunction we compare it to the correct seniority zero sector of FCI (FCI${}_0$). If the wavefunction is similar to the exact result, DOCI may provide a solid foundation for more advanced theories to treat both weak and strong correlation. We therefore extract the seniority zero sector of the FCI wavefunction by projecting out all elements with unpaired excitations,
\begin{equation}
 |FCI_0\rangle = \frac{P_{\Omega=0}|FCI\rangle}{\sqrt{\langle FCI |P_{\Omega=0}|FCI\rangle}}.
\end{equation}
We then seek to compare this wavefunction with the results from DOCI. The natural basis in which to examine the properties of the DOCI wavefunction is the orbitally optimised DOCI basis. This is found by variationally minimizing DOCI with respect to the orbitals. In practice this is done via orbital optimization of pCCD (oo-pCCD)~\cite{Stein14}. Due to the observed equivalence of the methods, this produces the optimal basis and the lowest possible DOCI energy.

\subsection{DOCI optimised basis}

We will begin by looking at a system where there are both weak and strong correlations. This will give us a sense of how well the DOCI wavefunction behaves in the most difficult case. In Figure~\ref{fig:ov_pCCD_PBC} we show the overlap and wavefunction energies of the DOCI and FCI${}_0$ wavefunction for a periodic $2\times 4$ Hubbard lattice with six particles. In the strong interaction limit, we see that the overlap between the two wavefunctions begins to decrease. While there is only a small reduction in the overlap, this results in a large change in the energies of the two wavefunctions. 

While we see that the small errors in the wavefunction can lead to a large change in the energy, it is also important to consider how DOCI performs in less extreme cases and later compare to a non-optimal basis. In Figure~\ref{fig:ov_pCCD}, we show the same overlap and energies of the DOCI and FCI${}_0$ wavefunction of a set of non-periodic systems both at half-filling and doped. We use non-periodic boundaries in order to make a direct comparison later where this will be important. We can see that the DOCI almost exactly reproduces the FCI${}_0$ wavefunction for the half-filled cases even though the system breaks seniority. We still see some difference in the energies for the doped system as was observed earlier.

\begin{figure*}[ht!]
\includegraphics[width=0.49\textwidth]{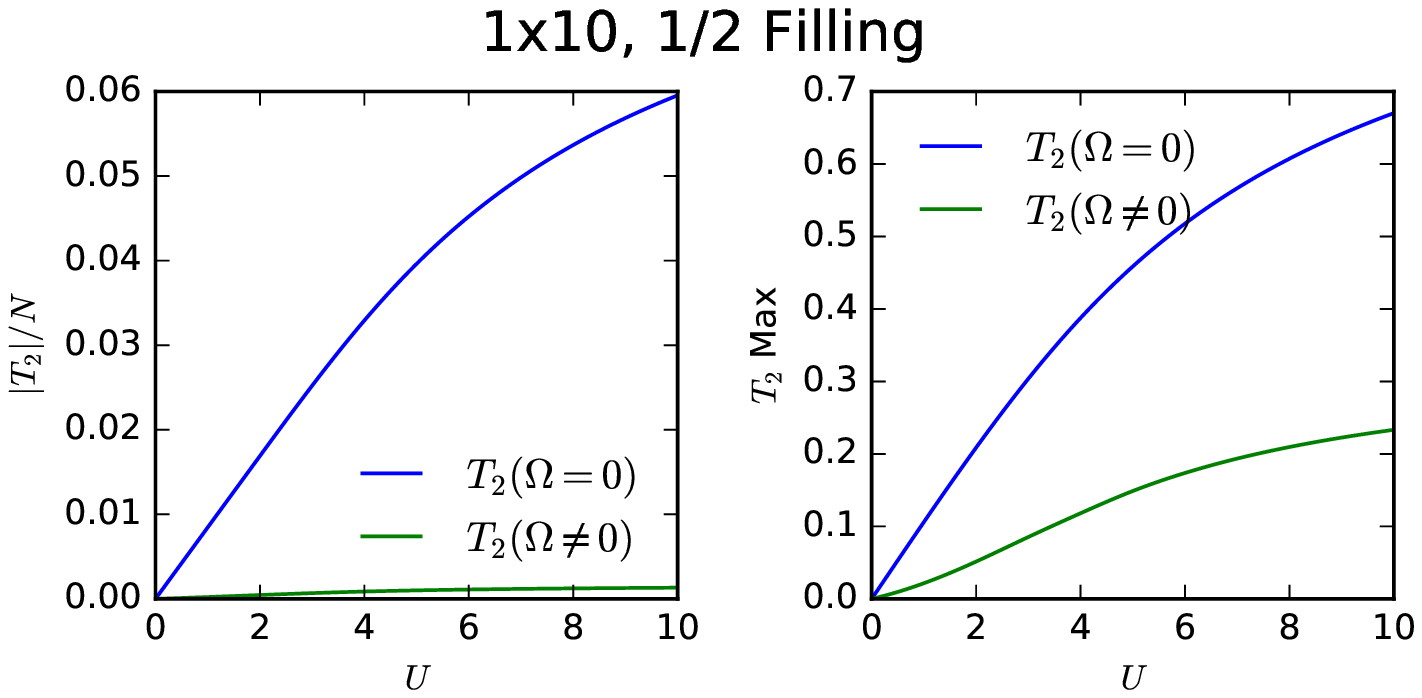}
\includegraphics[width=0.49\textwidth]{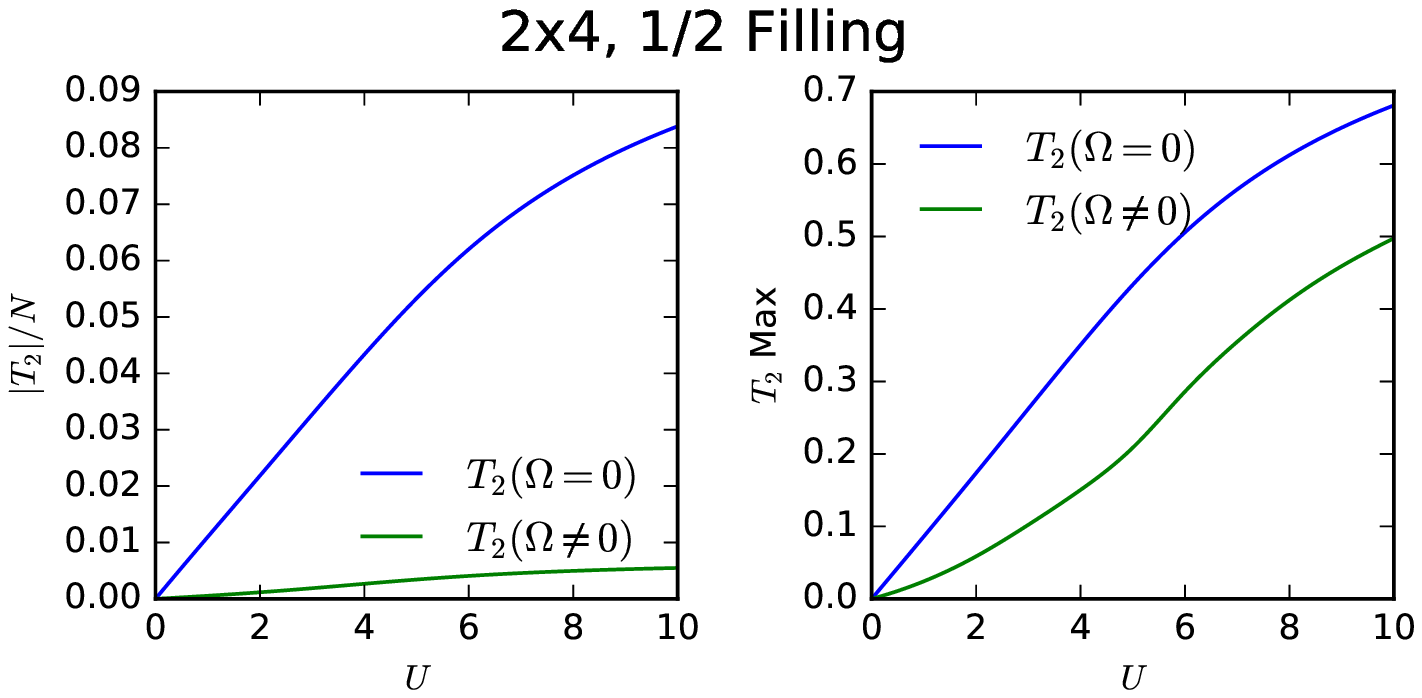}\\
\includegraphics[width=0.49\textwidth]{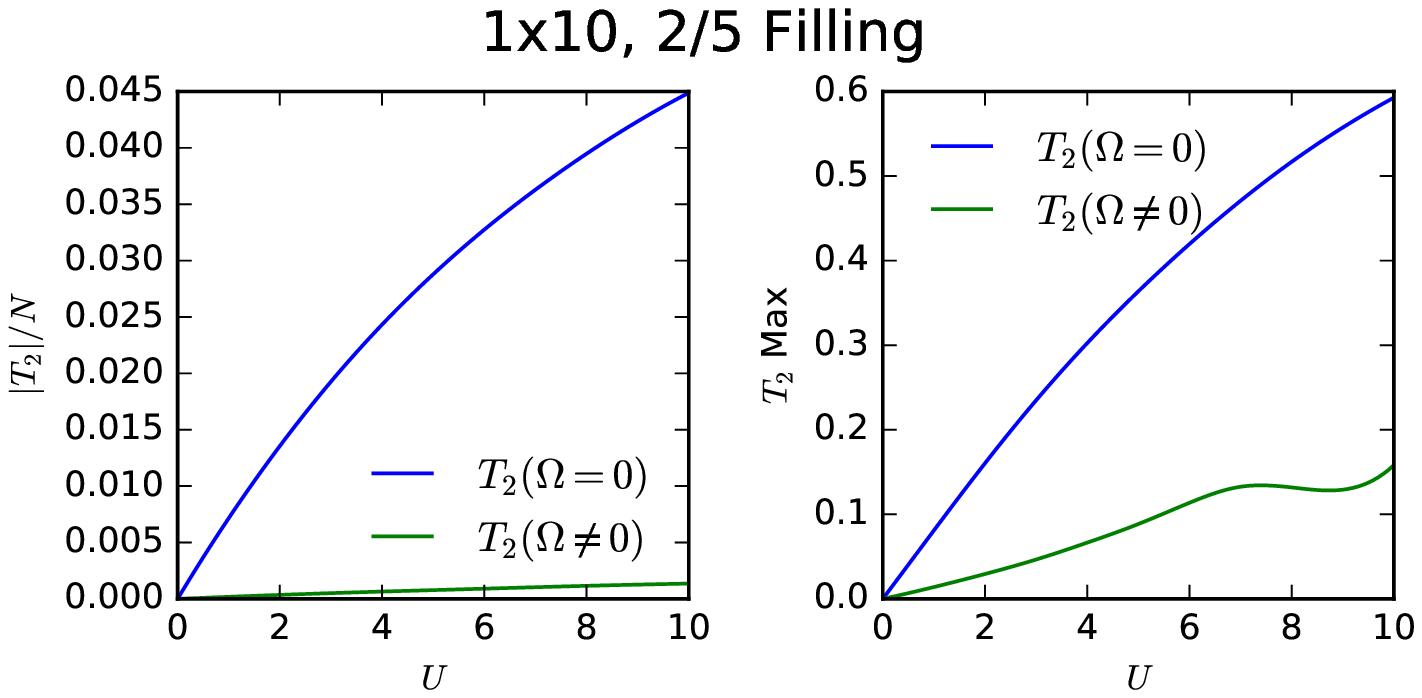}
\includegraphics[width=0.49\textwidth]{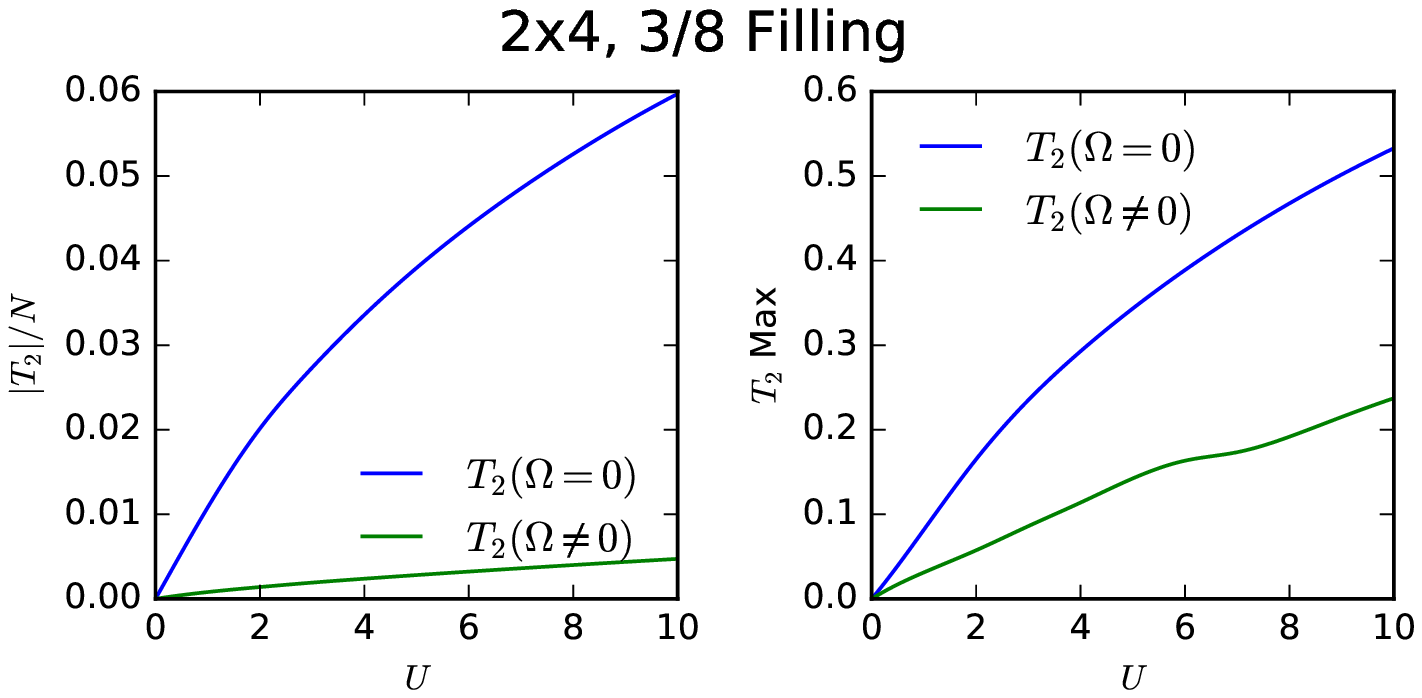}
\caption{\label{fig:T2size_pCCD} The average size and maximum values of the seniority zero and non-zero $T_2$ amplitudes in the oo-pCCD basis extracted from exact diagonalization.}
\end{figure*}

If we look at the relative size of the seniority sectors for the exact $T_2$ amplitudes extracted from FCI, we can compare the relative importance of the seniority zero wavefunction to the exact solution. These are important as most coupled cluster methods are truncated at double excitations and these are the highest order needed for the calculation of the energy. In Figure~\ref{fig:T2size_pCCD} we show the average size and maximum values of the seniority zero sector of $T_2$ ($t^{aa}_{ii}$) compared to the remaining amplitudes. It is clear that the seniority zero amplitudes are significant with smaller influence from the remaining amplitudes, particularly for the chains. This means that by getting the correct DOCI for these systems, we can accurately parameterise much of the exact $T_2$ with a mean-field cost method. This is important if we want to use this method to build a more advanced coupled cluster ansatz. We must keep in mind, however that this cannot be relied upon in cases of very strong interactions as demonstrated earlier (Figure~\ref{fig:ov_pCCD_PBC}).

\subsection{RHF basis}

While we have shown that in the optimal basis the DOCI wavefunction has strong overlap with FCI${}_0$ but has some deviation with fringe cases, we should compare this to a commonly used basis. While it has been observed that pCCD accurately approximates DOCI in both the RHF and optimised bases~\cite{Shepherd16}, the relation between DOCI and FCI${}_0$ can naturally be expected to be basis dependent. As seniority is a basis-dependent quantity, the coupling of the different seniority sectors through the Hamiltonian will also be basis dependent. Therefore, a less optimal basis can be expected to have greater coupling and greater renormalization of the FCI${}_0$ wavefunction through seniority breaking terms. 

\begin{figure*}[ht!]
\includegraphics[width=0.99\textwidth]{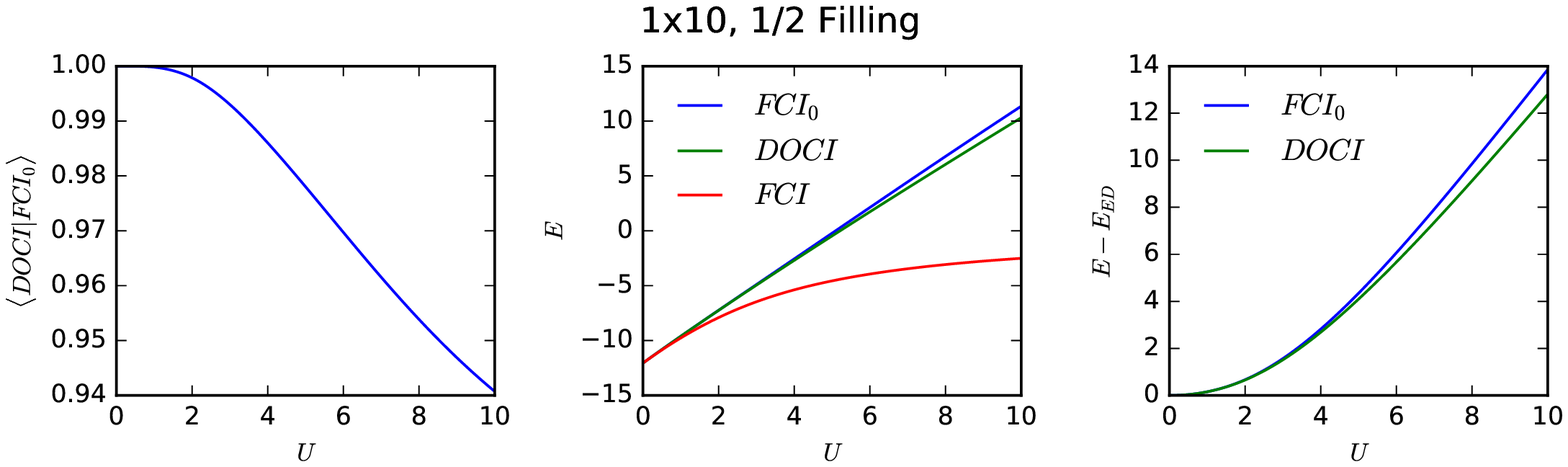}\\
\includegraphics[width=0.99\textwidth]{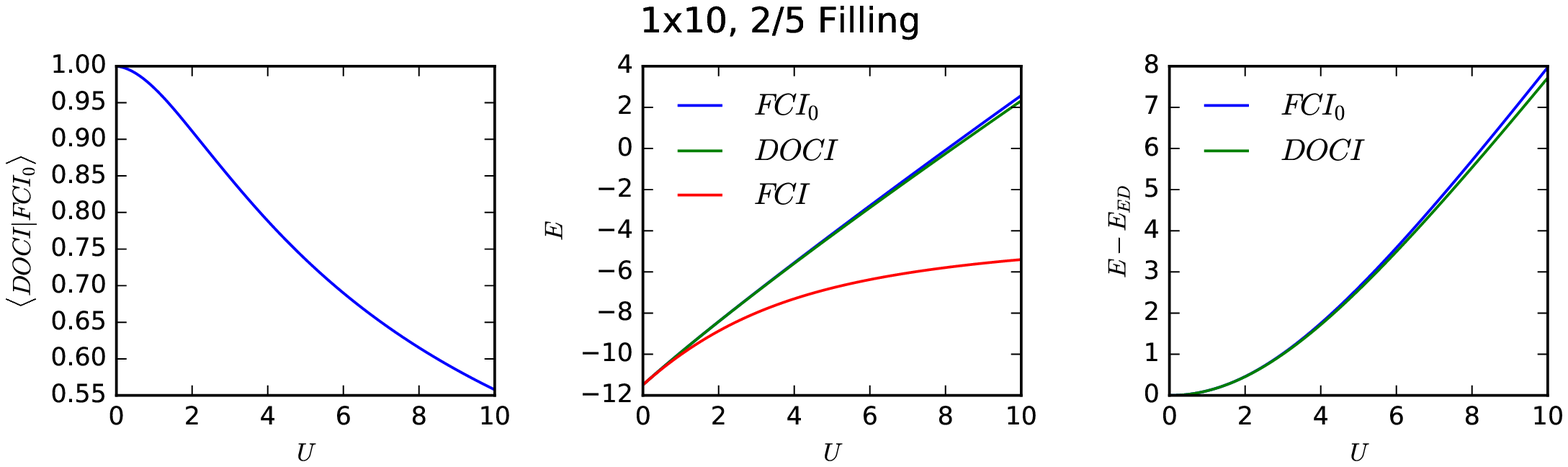}\\
\includegraphics[width=0.99\textwidth]{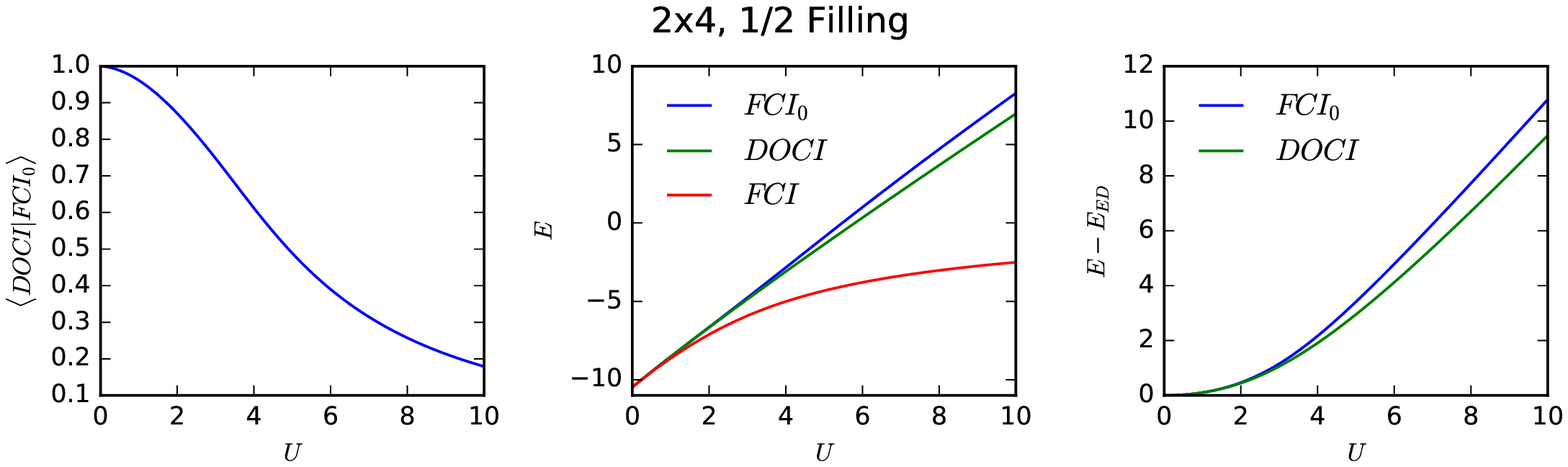}\\
\includegraphics[width=0.99\textwidth]{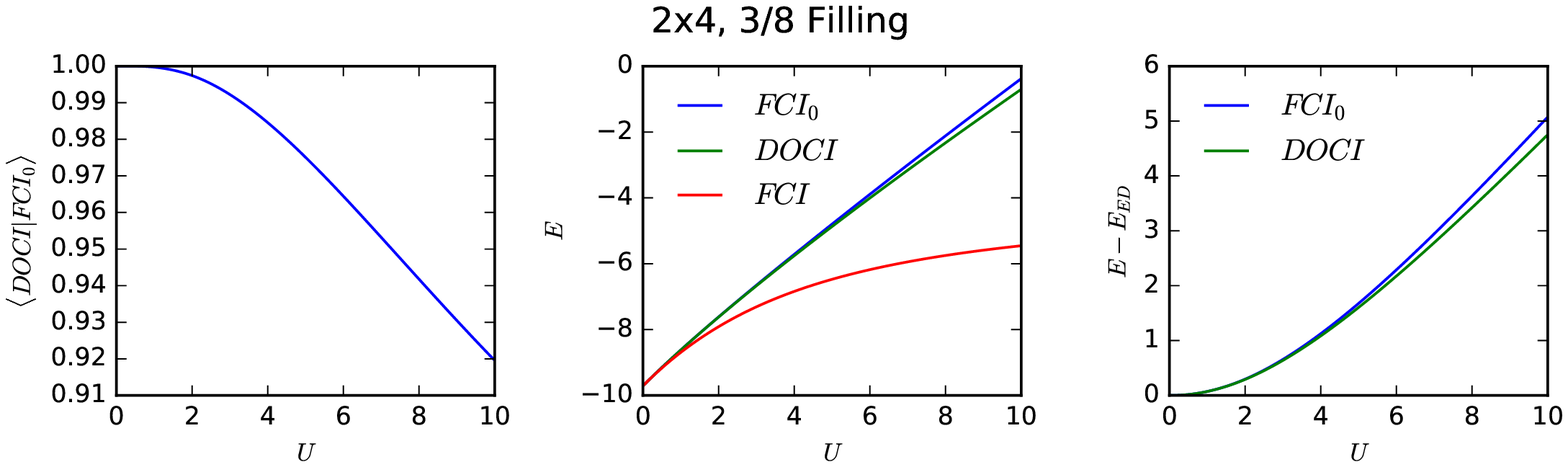}
\caption{\label{fig:ov_RHF} The overlap and energies of the DOCI and FCI${}_0$ wavefunctions for Hubbard chains and ladders in the RHF basis.}
\end{figure*}

This change in the FCI${}_0$ wavefunction is demonstrated in Figure~\ref{fig:ov_RHF} where we show the DOCI-FCI${}_0$ overlap and energies as before but in the RHF basis. The use of non-periodic systems is important here as the periodic systems have degeneracies in the single particle energies. In order to maintain a consistent basis and resulting definition of seniority, we use the non-periodic systems to lift this degeneracy. If we compare these results to those for the same systems in Figure~\ref{fig:ov_pCCD}, we immediately see a much greater disagreement between the two wavefunctions. The overlaps are generally smaller, and the differences in the energies larger. The differences in the energies is somewhat hidden by the scale of the overall energy error. This makes it clear that the pair-optimised basis used previously significantly improves the approximation of the DOCI wavefunction. The poorer quality of the DOCI wavefunction in this basis is likely the reason that RHF based fpCCD fails for stronger correlations~\cite{Stein14}. The smaller overlap for the doped chains compared to the ladders is likely due to the smaller doping fraction. Doped systems closer to half-filling are typically more more difficult to accurately describe due to competition between the N\'{e}el state at half-filling and the delocalized electrons at low filling.

If we also look at the relative size of the seniority zero sector of the exact $T_2$ amplitudes as before, we see that the seniority zero sector is still a significant portion of the wavefunction. This indicates that the FCI${}_0$ wavefunction is heavily renormalised by couplings to higher seniority sectors through the Hamiltonian. Optimizing the basis dramatically reduces this, greatly improving the approximation.

\begin{figure*}[ht!]
\includegraphics[width=0.49\textwidth]{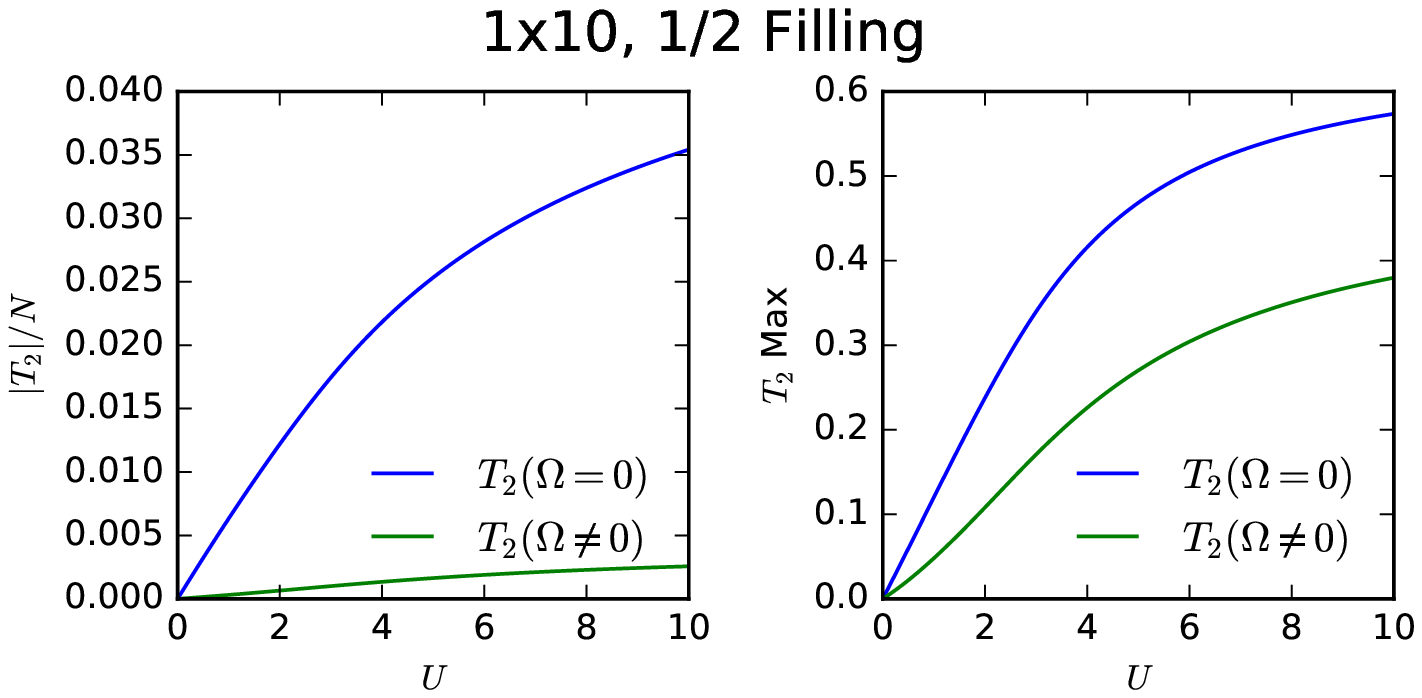}
\includegraphics[width=0.49\textwidth]{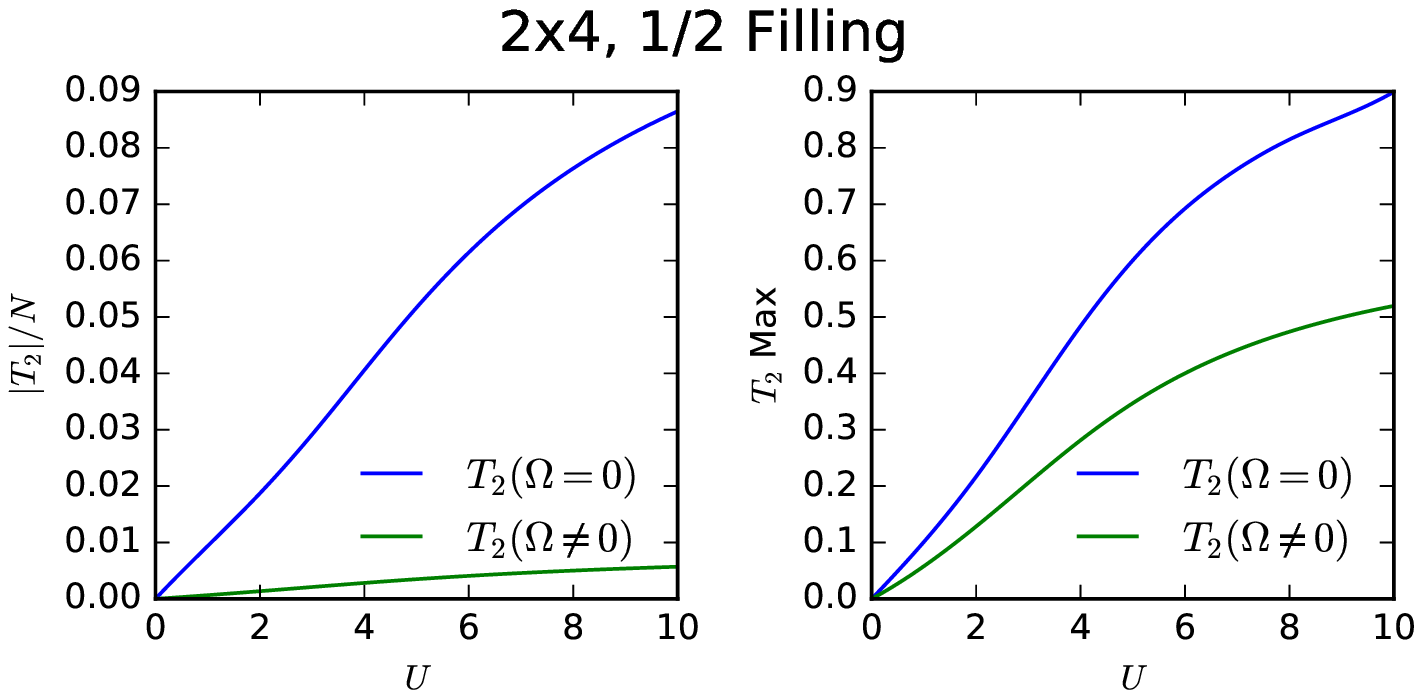}\\
\includegraphics[width=0.49\textwidth]{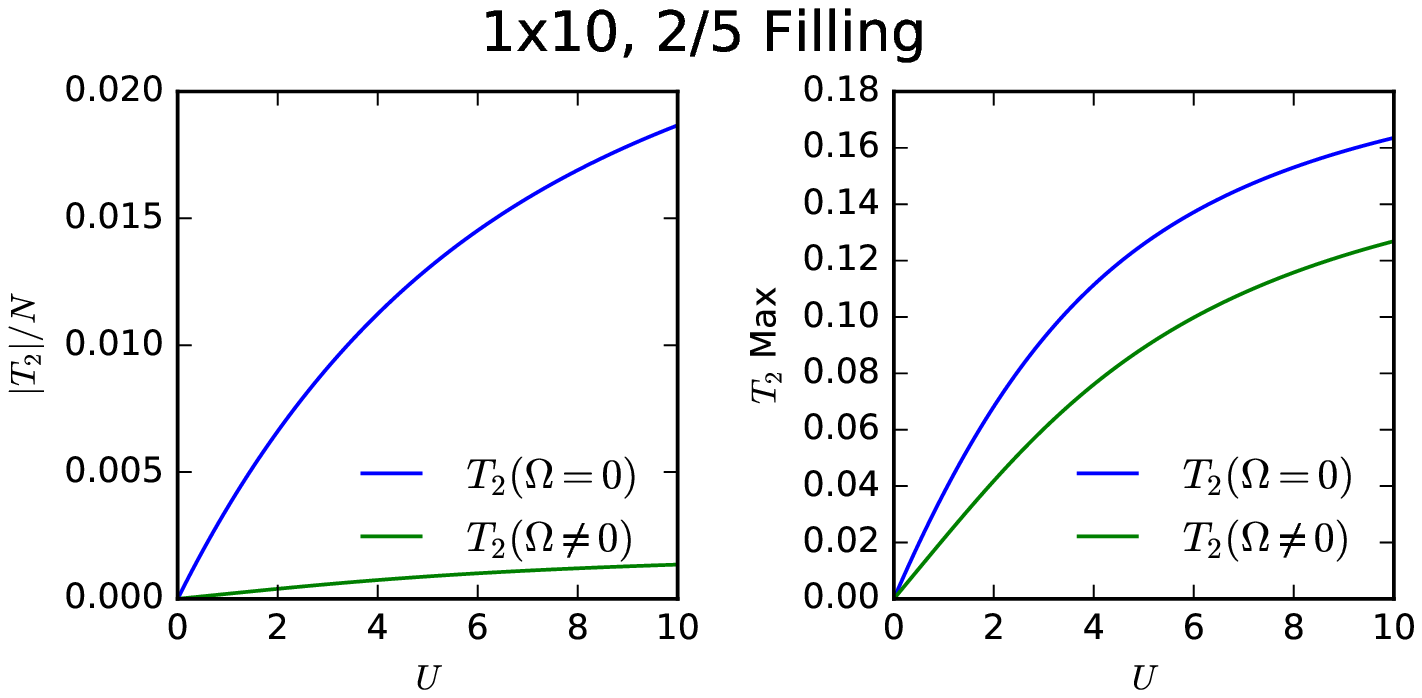}
\includegraphics[width=0.49\textwidth]{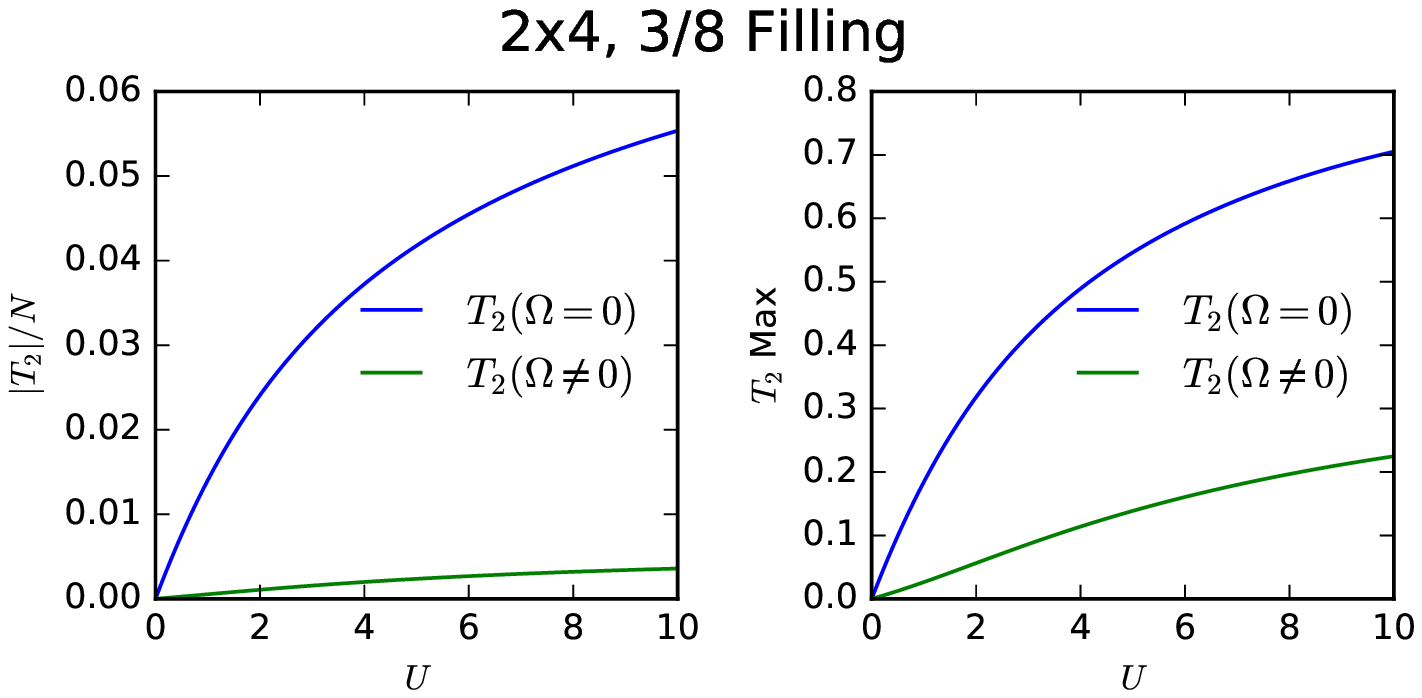}
\caption{\label{fig:T2size_RHF} The average size and maximum values of the seniority zero and non-zero $T_2$ amplitudes in the RHF basis extracted from exact diagonalization.}
\end{figure*}

\section{Conclusions}

It can be seen from these results that DOCI, and by extension pCCD, has some shortcomings when trying to accurately describe the seniority zero sector of the FCI wavefunction. While DOCI does reproduce the FCI${}_0$ wavefunction for half-filled Hubbard systems in the pair-optimised basis, care must be taken in some circumstances as it is not perfect, and small changes can be significant under strong correlations. Doped systems have some differences between the two wavefunctions, which has a strong effect on their respective energies for the strongly correlated case. In the RHF basis these differences are far more dramatic. The half-filled ladder has very poor agreement between the two wavefunctions. This is not surprising due to the failure of CCSD for this case. There is a larger difference between the two wavefunctions in all cases, but the energy difference is more difficult to see due to the large overall error for both wavefucntions.

While the differences are not dramatic in the pair-optimised basis, it is important to keep this in mind when treating strongly correlated systems. With large interactions, even small differences in the wavefunctions can dramatically affect the results. In addition, we cannot directly study larger systems due to the limitations of FCI calculations, so the behavior may change for larger systems and different filling fractions. With this in mind, it is still remarkable how much of the correct FCI${}_0$ wavefunction is recovered by DOCI in the proper basis. This may be a good point to start further approximations as long as it is understood that the DOCI wavefuncton is not perfect in every case. We will still need to consider possible ways to improve the results for the limiting cases where it does not work. Renormalization of the higher excitations such as in the symmetry projected wavefunctions~\cite{Qiu16} may be the correct answer for some systems, but this will be a topic of further study.

\section{Acknowledgements}

This work was supported by the U.S. National Science Foundation (CHE-1462434). 
G.E.S. is a Welch Foundation Chair (No. C-0036).

\bibliographystyle{tfo}
\bibliography{DOCI-FCI}

\end{document}